# VISIBILITY GRAPHS OF GROUND-LEVEL OZONE TIME SERIES: A MULTIFRACTAL ANALYSIS

AUTHORS: Carmona-Cabezas Rafael[1], Ariza-Villaverde Ana B.[1], Gutiérrez de Ravé Eduardo[1], Jiménez-Hornero Francisco J.[1]

[1]University of Córdoba (Spain)

CORRESPONDING AUTHOR: Carmona-Cabezas Rafael

VISIBILITY GRAPHS OF GROUND-LEVEL OZONE TIME SERIES: A MULTIFRACTAL ANALYSIS

AUTHORS: Carmona-Cabezas Rafael, Ariza-Villaverde Ana B., Gutiérrez de Ravé Eduardo, Jiménez-Hornero Francisco J.

## ABSTRACT

A recent method based on the concurrence of complex networks and multifractal analyses is applied for the first time to explore ground-level ozone behavior. Ozone time series are converted into 2-D complex networks for their posterior analysis. The searched purpose is to check the suitability of this transformation and to see whether some features of these complex networks could constitute a preliminary analysis before the more thorough multifractal formalism.

Results show effectively that the exposed transformation stores the original information about the ozone dynamics and gives meaningful knowledge about the time series. Based on these results, the multifractal analysis of the complex networks is performed. Looking at the physical meaning of the multifractal properties (such as fractal dimensions and singularity spectrum), a relationship between those and the degree distribution of the complex networks is found.

In addition to all the promising results, this novel connection between time series and complex networks can deal with both stationary and non-stationary time series, overcoming one of the main limitations of multifractal analysis. Therefore, this technique can be regarded as an alternative to give supplementary information within the study of complex signals.

Abbreviations: VG (Visibility graph), SBA (Sandbox Algorithm)

1. INTRODUCTION

Many studies have been performed about ground-level ozone over the last decades. The importance of ozone characterization and analysis lies on the fact that it is one of the main photochemical oxidants (due to its abundance). This irritant gas has serious repercussions for human health and harvests when its concentration is high (Doherty et al., 2009). Those exposed damages have an impact from the economical point of view, and according to Miao et al. (2017), they lead every year to losses of several billions of dollars.

Ozone is a secondary pollutant whose chemical formation and destruction mechanisms are known to be photochemical and nonlinear processes (Graedel and Crutzen, 1993; Trainer et al, 2000). These mechanisms highly depend on meteorological variables such as temperature, wind direction and mainly solar radiation (Graedel and Crutzen, 1993; Guicherit and Van Dop 1977), as it has been studied for the Mediteranean area in several works (Güsten et al., 1994; Kouvarakis et al., 2000; Ribas y Peñuelas 2004). In addition to that, ozone concentration depends as well on chemical precursors, such as nitrogen oxides and volatile organic compounds derived from the urban and industrial activity (Sillman, 1999). All these factors make the analysis of the temporal evolution of ozone a very complex task indeed.

Due to the facts exposed above, ozone studies based on traditional statistical analysis may provide a limited description of more complex dynamics of time series where the variability is high. The reason for this limitation is that these methods approximate and smooth the signal by means of fitting to functions, with the derived loss of information. Besides, they base their results on one (time) scale, while the physical phenomenon can appear at several scales due to the number of variables in play (Zeleke and Si 2006). On the contrary, multifractal methods can be used to save this drawback, since they work directly with the raw data extracting the information from their singularities. Furthermore, fractals (and multifractals) are characterized for being self-similar when divided into smaller parts (i.e. they are scale-independent) or at least their statistical properties (Mandelbrot 1982). That way, if a natural phenomenon can be characterized by means of multifractal parameters, these will be able to describe it for a range of scales.

In the presented work, a link between the multifractal analysis and complex networks has been tested for the description of ground-level ozone dynamics. To that purpose, ozone concentration time series have been transformed into 2-D visibility graphs (VG) (whose topology inherits the features of the associated time series) and then evaluated using two methods for multifractal analysis: the sandbox method in order to compute the generalized fractal dimensions (Rényi spectrum) and the approach introduced by Chhabra and Jensen (1989) for the calculation of the α-spectrum as an independent value from the other.

With this study, the main purpose is to check the suitability of the multifractal analysis performed over the VGs by connecting their degree distribution with Rényi and α-spectra. It could be expected, since the resulting graph stores indeed much of the original information and properties of the original time series (Lacasa et al., 2008).

2. <u>MATERIALS AND METHODS</u>

2.1. <u>Data</u>

The information that has been used in the analysis of this work corresponds to a 10-min ozone concentration data collected for the months of January, April, July and October in 2007. These time series of ozone concentration can be seen in Figure 1, for the four months analyzed in this study.

The chosen region is the western part of Andalusia (Spain) since as exposed by Domínguez-López et al. (2014), this area meets the weather conditions (high temperatures and solar radiation), orographic (the valley of the Guadalquivir river) and anthropic (four capitals and two important industrial centers such as the chemical pole of Huelva and the Bay of Algeciras) to be potentially vulnerable to pollution by surface-level ozone. The measurements were performed at the urban station located in Lepanto, Córdoba (37.53° N, 4.47° W). The cited station belongs to the regional network in charge of controlling the air quality in Andalusia, co-finanzed by the Consejería de Medioambiente (Regional Environmental Department) and the European Union. This station is located at 117 m of altitude and the average temperature for each moth is 8, 15, 28 and 18 °C (January, April, July and October, respectively). The average direct solar radiation is 310.2, 577.8, 991.4 and 419.3 MJ/m$^2$ (again for the four months ordered).

The climate of the zone of study, according to the Köppen-Geiger classification, is defined as Csa with warm average temperatures and hot and dry summer.

As it can be seen in Figure 1, the ozone concentrations are especially high in summer (July) and low in Winter (January). As it has been previously discussed for this region by Adame et al. (2007) and Jiménez-Hornero et al. (2010a), that is due the fact that the conditions for the ozone creation are more suitable from the end of January. The progressive raise of temperature and solar radiation that reaches its peak in July allows those higher creation rates and thus its concentration. One of the reactions that governs the ozone production can be found below (Graedel and Crutzen, 1993).

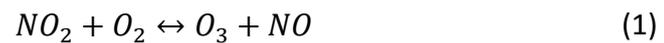

$$NO_2 + O_2 \leftrightarrow O_3 + NO \qquad\qquad (1)$$

It is a reversible photochemical reaction which tends to the ozone production when there is energy available in form of light (right sense of the arrow) and in the other way when there is not. For that reason, the higher values of ozone are always found during the day and vice versa, happening the same with summer and winter respectively, as discussed before.

## 2.2. Visibility graph

One of the main fields of application of the multifractal analysis is referred to the study of graphs and complex networks. A graph can be defined as a set of vertices, points or nodes connected to each other by lines that are usually called *edges*. A tool to transform time series into a graph was presented by Lacasa et al. (2008). This new complex network receives the name of *visibility graph* and has been proven to inherit many of the properties of the original series (being some multifractal properties among them).

In order construct the visibility matrix which contains the information of all the nodes of the new system, it is necessary to stablish a criterion to discern whether two points would be connected or not. This criterion reads as follows: two arbitrary data from the time series $(t_a, y_a)$ and $(t_b, y_b)$ have visibility (and would become two connected nodes in the graph) if any other data point $(t_c, y_c)$ between them $(t_a < t_c < t_b)$ fulfills the following condition:

$$y_c < y_a + (y_b - y_a)\frac{t_c - t_a}{t_b - t_a} \qquad (2)$$

In Figure 2, an application of this method for a simple time series is given as illustration. As it can be seen, the original time series has been transformed into a complex network. The complexity of the original series is inherited by the new graph, as it has been found by Lacasa et al. (2008, 2010), meaning that for instance a periodic time series, would lead to a regular graph.

The result of applying this visibility method is a NxN adjacency binary matrix, being N the number of points in the set. Each row of the matrix contains the information of a different node, such as $a_{ij} = 1$ means that the node $i$ and $j$ have visibility; whereas $a_{ij} = 0$ means that there is no edge between them. The resulting matrix has several properties that can be used to simplify the algorithm and thus reduce the computational required time.

- Hollow matrix: All the elements in the diagonal are zero ($a_{ii} = 0$), because there is no visibility of an element with itself, since there are no intermediate points to fulfills the criterion.

- Symmetric matrix: The elements satisfy $a_{ij} = a_{ji}$, due to the reciprocity of the visibility between two nodes.

- Nearest neighbors: Because each point always sees the closest previous and next node, the elements surrounding the diagonal are always 1 ($a_{ij} = 1$ for $j = i \pm 1$).

Taking all of that into account, the visibility matrix A has a general form as shown below:

$$A = \begin{pmatrix} 0 & 1 & \dots & a_{1,N} \\ 1 & 0 & 1 & \vdots \\ \vdots & 1 & \ddots & 1 \\ a_{N,1} & \dots & 1 & 0 \end{pmatrix}$$

The degree of a node ($k_i$) can be defined as the number nodes that have reciprocal visibility with the first one ($k_i = \sum_j a_{ij}$). In Figure 2, the degree of the first node is $k = 3$, for the second one $k = 2$, for the third one $k = 3$, and so on. From the degree of each one of the nodes present in the VG, it is possible to obtain the degree distribution of the sample ($P(k)$), which is nothing but the probability that every degree has.

Previous works have shown how the analysis of this degree distribution built from the VG can effectively describe the nature of the time series (Lacasa et al., 2008; Mali et al., 2018), distinguishing between periodic, random or fractal series for instance. Therefore, by studying the degree distribution, one can get a first insight on the behavior of the ozone time series before stepping into a more complex multifractal analysis. As stated by Lacasa et al. (2009, 2010), time-series that have VGs with degree distributions that follow power laws such as $P(k) \propto k^{-\gamma}$, can be considered as scale free.

### 2.3. Multifractal measurements

While fractal analysis is based on the complexity of a fractal set, the multifractal approach can describe the distribution of a given measure over a fractal object (Mandelbrot, 1974; Halsey et al., 1986). It implies the possibility of having different densities depending on the region of application.

There are two ways of representing multifractals: the generalized fractal dimensions $D_q$ and the singularity or multifractal spectrum ($f(\alpha)$). Both of them are discussed below separately. Typically, the multifractal analysis has been widely performed by means of the fixed-size algorithms (FSA), that rely on the subdivision of the system into smaller parts with equal size and then that size is changed iteratively. The method used in this work is the "sand-box algorithm" and will be discussed later on.

As stated before, the first of the measurements when it comes to multifractal analysis are the generalized fractal or Rényi dimensions $D_q$, which describe the scaling exponents of the $qth$ moments of the system and can be defined (Feder, 1988) as:

$$D_q = \frac{1}{q-1} \lim_{\delta \to 0} \frac{\ln Z_q(\delta)}{\ln \delta} \quad \forall \; q \neq 1 \tag{3}$$

$$D_1 = \lim_{\delta \to 0} \frac{\sum_{i=1}^{N_c(\delta)} \mu_i(\delta) \ln \mu_i(\delta)}{\ln \delta} \qquad (4)$$

With $q$ the moment order, $\delta$ the size of the used cells to cover the system, $Z_q(\delta)$ the partition function, $N_c(\delta)$ the number of cells with length $\delta$ and $\mu_i(\delta)$ the probability measurement of each cell. The expression for $D_1$ is obtained taking the limit of $D_q$ when $q \to 1$.

From these generalized fractal dimensions, it must be pointed out that $D_{q=0}$ corresponds to the fractal dimension of the given system or *box-counting dimension*, $D_{q=1}$ is the so-called *information entropy* and $D_{q=2}$ the *correlation dimension.* The limits of $D_q$ when $q$ goes to $-\infty$ and $+\infty$ describe the scaling properties of the regions where the measure is more rarified and concentrated, respectively.

The other set of multifractal parameters is the so-called singularity spectrum ($f(\alpha)$), as commented previously. A frequent method to determine is based on the use of a Legendre transform from mass exponents $\tau(q)$. However, some authors as Chhabra and Jensen (1989) and Veneziano et al (1995), state that it can lead to some errors due to the inclusion of spurious points and error amplification from the derivative. Furthermore, it does not yield independent measurements from the Rényi spectrum, as $\tau(q) = (1 - q)D_q$. As an alternative to this and more focused on experimental data, Chhabra and Jensen (1989) proposed a direct method to determine the α-spectrum, overcoming the drawbacks referred before. This technique relies on the normalized measure $\beta_i(q)$, $\mu_i$ in the original work, for computing the probabilities of the boxes of radius $r$:

$$\beta_i(q,r) = [P_i(r)]^q / \sum_j [P_j(r)]^q \qquad (5)$$

With $P_i(r)$ the different fractal measurements for each box of radius $r$ (number of nodes in this case). From this, $f(\alpha)$ and $\alpha$ can be retrieved using the next expressions:

$$f(q) = \lim_{r \to 0} \frac{\sum_i \beta_i(q,r) \log[\beta_i(q,r)]}{\log r} \qquad (6)$$

$$\alpha(q) = \lim_{r \to 0} \frac{\sum_i \beta_i(q,r) \log[P_i(r)]}{\log r} \qquad (7)$$

Being $\alpha$ known as the Lipschitz-Hölder exponent. In practice, those quantities are computed from the slope of $\sum_i \beta_i(q,r) \log[\beta_i(q,r)]$ over $\log r$ for $f(q)$; and $\sum_i \beta_i(q,r) \log[P_i(r)]$ over $\log r$ for $\alpha(q)$. This slope is determined by means of a linear regression in the same range of radii where the other fractal measures are computed.

## 2.4. Sandbox algorithm

The sandbox algorithm (SBA) was firstly introduced by Tél et al. (1989) and developed by Vicsek et al. (1990), as an improvement of the previously used fixed-size box-counting methods for computing the generalized fractal dimensions. The main advantage of this method with respect to other box-counting FSA is that it is capable of properly determine the side corresponding to the negative values of $q$ from both the Rényi and singularity spectra.

The basic idea behind the SBA is that for each radius, a number of randomly placed boxes are selected, and they are always centered in a non-zero point of the system (a node). In that way the entire network is covered with those boxes by choosing a sufficiently high number of them. For each box (B), the probability measurement used is computed as shown in equation (8).

$$\mu(B) = \frac{M(B)}{M_0} \tag{8}$$

Once that quantity is computed for each sandbox for a given radius, the generalized fractal dimensions can be obtained as explained previously in the multifractal measurement part. Applying it to equation (3), the following formula is obtained for $D_q^{sb}$:

$$D_q^{sb} = \frac{1}{q-1} \lim_{r \to 0} \frac{\ln\langle \mu(B)^{q-1} \rangle}{\ln r} \quad \forall \ q \neq 1 \tag{9}$$

And the expression when $q = 1$ for the SBA also can be adapted from equation (4):

$$D_1^{sb} = \lim_{r \to 0} \frac{\langle \ln \mu(B) \rangle}{\ln r} \tag{10}$$

With the aim of implementing the SBA on the VG, the steps to follow are:

1) The original time series is transformed into a VG, resulting on a matrix as described in the previous section.

2) A range of different radii is set in order to cover the entire network. These radii are chosen between 1 and N, being N the total number of nodes. For this case, the typical values are $r \in [1, 100]$, because larger values have proven to give the same information, since the curves start to saturate.

3) Then, $N_c$ centers are selected for each radius. This number is inversely proportional to the radius itself, because the greater the box is, the smaller number of boxes will be needed to cover the network. The location of the centers is randomly chosen within the nodes of the graph.

4) For each sandbox, the amount of nodes inside the box connected to the center ($M$) are counted, giving $N_c$ values of $M(B)$. From here, the quantity $\mu(B)$ (for every box) is computed by means of dividing the previous $M(B)$ by the total amount of nodes connected to the center $M_0$. Then $\mu(B)$ is used to compute the partition function for all the $q$ values chosen, being the average in equations (9) and (10) over all the sandboxes created.

5) The steps 3) and 4) are then repeated for all the radii considered, obtaining a value of the partition function for each one of them and $q$.

## 3. RESULTS AND DISCUSSION

### 3.1. Degree distribution

Firstly, a fast method to check the fractal nature of the ozone concentration time series, before performing a deeper study, is described here. This method consists on analyzing the degree distribution of the VGs.

The mentioned distribution is computed as the number of nodes that have a given degree and divided by the total number of them in the VG. It is clear that the greater the degree is, the less likely to be repeated within the network it will be; because large degrees are exclusive of nodes with the highest concentration (which we will refer as *hub*) due to their typically high visibility. As those hubs correspond normally to the extreme values of ozone concentration of each day, their likeliness will be lower, since the most repeated ones will be the ones close to the average.

That leads to an expected distribution with a negative trend, as can be seen in Figure 3.

Once the degree distributions of the different months are computed, a clear fractal behavior is observed since all of them follow a power law $P(k) \propto k^{-\gamma}$ as expected, with a linear part in the last part of the log-log plot. After computing the slope for the values of $k \geq 30$ in all the cases, the $\gamma$ parameter can be determined, obtaining that the biggest and the lowest ones correspond to July and January respectively. Although this is a fast and direct method to determine whether the series is fractal or not, it does not give much detailed information about the differences between each moth, since all of them share a very similar degree distribution. Hence a deeper analysis devoted to the multifractal properties of the series has been done with that aim, as will be presented in the following section, in order to give some light to the usability of this degree distribution and some of its parameters.

## 3.2. Multifractality

In this part, the authors present two independent methods to study the multifractality of the signal, both previously described: the generalized fractal (Rényi) dimensions and the $\alpha$ spectra.

For the analysis of the Rényi dimensions of the samples, the quantities $\ln\langle\mu(B)^{q-1}\rangle / (q-1)$ for $q \neq 1$ and $\langle\ln\mu(B)\rangle$ for $q = 1$ are plotted against $\ln r$, as seen in Figure 4. The interval used for the radii goes from $r = 1$ to $r = 100$, as it was proven to be enough in order to reproduce the expected linear behavior (after several tests). It must be pointed out that the interval taken for the linear regression is always selected so that the Pearson correlation coefficient is $r \geq 0.99$. For April, July and October, the chosen range has been from $\ln r = 3$ up to end of the curve; whereas for January the range had to be taken such as $\ln r \in [2, 4]$ because the linear part is larger and to avoid an artifact that appears for higher radii and made the results misleading (can be observed in Figure 7). In all those cases, the values correspond with the higher radii used in the SBA, in accordance with previous studies that used the same technique (Ariza-Villaverde et al., 2015; De Bartolo et al., 2004; Jiménez-Hornero et al. 2013). After the linear regressions are performed for all values of $q$ of each month, the Rényi spectrum can be constructed following equations (9) and (10). In Figure 5 the result of such procedure is shown in the form of the generalized fractal dimensions. After observing the graph, it can be inferred that the VG properly reproduces the multifractal behavior of the series, which was previously demonstrated to be so (He, 2017; Jiménez-Hornero et al., 2010a, 2010b; Pavón-Domínguez et al., 2013), since for all the studied months $D_0 > D_1 > D_2$ (see Table 1). The difference between the maximum value of $D_q$ and the minimum ($\Delta D_q$) is usually taken as a measure of the multifractality degree of the signal (ozone concentration in this case) (Ariza-Villaverde, 2013; Telesca et al., 2004). Looking at Table 1, this degree seems to be sensibly higher for July than the rest of the months, followed by April and October (with a very similar value) and then the last one is January.

Now, the fractal dimensions themselves from the Rényi spectrum ($D_0$, $D_1$ and $D_2$), are being discussed. Firstly, the capacity dimension or "box-counting" dimension $D_0$ is related with the number of boxes needed in order to cover the fractal object. The more boxes are needed, the more extended is the fractal object and then the grater would be $D_0$. Since in this case the fractals are different for each iteration (each node sees a different visibility network), the final behavior would be expected to be an average of the whole system. In the context of these complex networks, a high extension of the object would mean a bigger degree (then related to hubs). Following that logic, a higher value of $D_0$ is related with greater number of hubs in the system, meaning that one could expect the average degree $\bar{k}$ of the distribution increased. This is found in the results as can be observed in Table 1. Values for the months of January and April are very close to each other, while the one for July is sensibly the highest and October has an intermediate value. This result is in accordance to the behavior of the ozone concentration, since for instance the month of July is the one with the greatest temperature and UV radiation rates and therefore elevated values are reached more often, producing a major number of hubs in the VG.

The second parameter is the entropy dimension $D_1$, which is usually known as the dimension that is related to the uniformity of the data and how different is the probability of certain regions to be visited by a randomly chosen box with respect to others. A way of measuring its uniformity is looking at the difference between $D_0$ and $D_1$, because when they are equal it means that the sample is uniform. The greater $(D_0 - D_1)$, the less uniform it is. Also, another reason to take the difference instead of the absolute value is because we are interested on comparing the curves and their behavior, and since the $D_0$ of each one is different, it is necessary to stablish a reference point. When it is translated to the context of VG, a greater uniformity would mean less difference between the degrees of the sample, hence the standard deviation of the degree would be decreased. In that case, what should be observed in the original ozone data is that the differences encountered in the data are less significative (what is indeed observed for the months with lower ($D_0 - D_1$). The most uniform would be January and again the other extreme is found for July. The reason for this is that in July the differences encountered between day and night concentrations of ozone are much greater than the other months, since during the night they remain at their minimum values for all months. In contrast, the maximum values are much higher in summer than the rest of the year because of the higher UV exposures and the opposite case for winter.

The last parameter that has been computed from the Rényi spectrum is the correlation dimension $D_2$. For the same objective reason as in ($D_0 - D_1$), here ($D_0 - D_2$) values are discussed as well. Once again, January and July exhibit the extreme cases whereas the other two months are located in between. In this case the authors have not been able to directly relate this magnitude with any property of the degree distribution of the VGs as for ($D_0 - D_1$) and $D_0$.

As commented before, another important feature that can be studied is the spectrum of $\alpha$, following the equations (6) and (7). Applying a similar approach to the one used to compute $D_q$, a linear regression is performed for each value of $q$ to compute the corresponding magnitude ($\alpha$ in Figure 6 and $f(\alpha)$ in Figure 7). For each month, the range for the regression has been chosen to be the same that was used for the Rényi dimensions, in order to conserve the scale used. In this case, the obtained curves have as well a positive trend that increases with $q$ for the case of the curves of Figure 6; and first increases and then decreases for Figure 7 (as expected, because $f(\alpha)$ should have a maximum and then drop). As commented before, the artifact that made us change the chosen interval for the regression in January can be observed in Figure 7 for high values of radius.

Once both $\alpha$ and $f(\alpha)$ are resolved for all the possible values of $q$, the spectrum of every month can be plotted, being the ones shown in Figure 8. All the spectra as can be seen have their maximum at $f(\alpha) = 1$ and start at $\alpha = 0$, while the overall shape depends on each case. Several properties of the underlying signal can be extracted from the spectra. The width of the curve $W$ is shown in Table 1, as well as the position of the curve maximum $\alpha_0$. As in the case of generalized fractal dimensions $D_q$, July is the month with the widest spectrum and highest $\alpha_0$. January as well exhibits the lowest values, whereas April and October have intermediate ones. This width $W$ is related with the multifractality degree of the signal (Ariza-Villaverde, 2013; Telesca et al., 2004).

According to the shapes of the curves, the spectra are strongly non-symmetric, being the right tail much more pronounced than the left one for all the different months. The right side is usually associated to the homogeneity of low values in a temporal distribution of data, whereas the left one is related with the same feature of the high values instead. Therefore, the behavior of the four months is very similar for high values (left tail) while it differs significantly for low ones (right tail). In Figure 8, all of them show an heterogenous behavior since the $\alpha$ distribution is not uniform along the curve, being concentrated on the left and right extremes. Nevertheless, it is possible to extract from that figure that this heterogeneity is more pronounced in July. This fact suggests an influence of the higher UV radiation that creates a greater difference between day and night ozone concentrations. This is in accordance with what was shown in the previous analysis described above for ($D_0 - D_1$).

4. <u>CONCLUSIONS</u>

The results of this work show that the multifractal analysis of VGs from ground-level ozone concentration time series is a suitable tool to describe the seasonal dynamics of this air pollutant. VGs have proven to have several advantages such as: i) their topology inherits the features of the associated time series, which ends up resulting on supplementary information through the degree distribution; ii) they can be use for both stationary and non-stationary time series, removing the multifractal analysis requirement of dealing only with stationary processes. Besides, VGs filter out trends in the signal, making unnecessary to apply detrended methods. iii) Also, they can be applied to multivariate time series, which can be very helpful in order to find correlations between tropospheric ozone and its precursors for instance; iv) and finally, this novel connection between time series and complex networks opens a broad range of possibilities within the study of complex signals.

When it comes to the multifractal analysis performed in this work, both the SBA and the Chhabra and Jensen method for the Rényi and singularity spectra respectively, where chosen based on their advantages with respect to other. The SBA overcomes the drawbacks of the box-counting algorithm for the computation of the generalized fractal dimensions for negative probability moment orders $q$. Furthermore, the Chhabra and Jensen method for the α-spectrum does not need a Legendre transform to be applied (with the associated error to it) and gives and independent measurement from the SBA as well, being more robust for comparison.

After applying the methods mentioned above, clear and coherent results that fits the expected behavior of the ozone dynamics were found for the different months. Furthermore, several properties can be directly inferred from the degree distribution almost at first sight, meaning a powerful tool for predicting results before any more complex data treatment is performed. The same relations between multifractal parameters such as capacity dimension, $W$ of the α-spectrum amongst others are found in simple statistical parameters as the average or standard deviation of the degree distribution of the VG.

To conclude, authors would like to point out that this promising technique could be extended to other applications due to the many possibilities that complex networks have. One possible target would be the relation between ozone and its precursors, which could be looked by means of relating the VG of each one separately or using multi-layer networks as proposed by Lacasa et al. (2015).

## 5. ACKNOWLEDGEMENTS

The FLAE approach for the sequence of authors is applied in this work. Authors gratefully acknowledge the support of the Andalusian Research Plan Group TEP-957. R. Carmona-Cabezas truly thanks the backing of the "Programa de Empleo Joven" (European Regional Development Fund / Andalusia Regional Government).

## 6. REFERENCES

Adame, J.A., Lozano, A., Bolívar, J.P., De la Morena, B.A., Contreras, J., Godoy, F., 2008.

Behavior, distribution and variability of surface ozone at an arid region in the south of Iberian Peninsula (Seville, Spain). Chemosphere 70, 841–849. https://doi.org/10.1016/j.chemosphere.2007.07.009

Ariza-Villaverde, A.B., Jiménez-Hornero, F.J., Gutiérrez de Ravé, E., 2015. Influence of DEM resolution on drainage network extraction: A multifractal analysis. Geomorphology 241, 243–254. https://doi.org/10.1016/j.geomorph.2015.03.040

Ariza-Villaverde, A.B., Jiménez-Hornero, F.J., Gutiérrez de Ravé, E., 2013. Multifractal analysis applied to the study of the accuracy of DEM-based stream derivation. Geomorphology 197, 85–95. https://doi.org/10.1016/j.geomorph.2013.04.040

Chhabra, A., Jensen, R.V., 1989. Direct determination of the f(α) singularity spectrum. Physical Review Letters 62, 1327–1330. https://doi.org/10.1103/PhysRevLett.62.1327

De Bartolo, S.G., Gaudio, R., Gabriele, S., 2004. Multifractal analysis of river networks: Sandbox approach. Water Resources Research 40. https://doi.org/10.1029/2003WR002760

Feder, J., 1988. Fractals, Physics of Solids and Liquids. Springer US.

Graedel, T.E., Crutzen, P.J., 1993. Atmospheric change: an earth system perspective. Atmospheric change: an earth system perspective.

Guicherit, R., van Dop, H., 1977. Photochemical production of ozone in Western Europe (1971–1975) and its relation to meteorology. Atmospheric Environment (1967) 11, 145–155. https://doi.org/10.1016/0004-6981(77)90219-0

Güsten, H., Heinrich, G., Weppner, J., Abdel-Aal, M.M., Abdel-Hay, F.A., Ramadan, A.B., Tawfik, F.S., Ahmed, D.M., Hassan, G.K.Y., Cvitaš, T., Jeftić, J., Klasinc, L., 1994. Ozone formation in the greater Cairo area. Science of The Total Environment 155, 285–295. https://doi.org/10.1016/0048-9697(94)90507-X

Halsey, T.C., Jensen, M.H., Kadanoff, L.P., Procaccia, I., Shraiman, B.I., 1986. Fractal measures and their singularities: The characterization of strange sets. Phys. Rev. A 33, 1141–1151. https://doi.org/10.1103/PhysRevA.33.1141

He, H., 2017. Multifractal analysis of interactive patterns between meteorological factors and pollutants in urban and rural areas. Atmospheric Environment 149, 47–54. https://doi.org/10.1016/j.atmosenv.2016.11.004

Jiménez-Hornero, F.J., Ariza-Villaverde, A.B., De Ravé, E.G., 2013. Multifractal

description of simulated flow velocity in idealised porous media by using the sandbox method. Fractals 21, 1350006. https://doi.org/10.1142/S0218348X13500060

Jiménez-Hornero, F. J., Gutiérrez de Ravé, E., Ariza-Villarverde, A.B., Giráldez, J.V., 2010a. Description of the seasonal pattern in ozone concentration time series by using the strange attractor multifractal formalism. Environmental Monitoring and Assessment 160, 229–236. https://doi.org/10.1007/s10661-008-0690-y

Jiménez-Hornero, Francisco J., Jiménez-Hornero, J.E., Gutiérrez de Ravé, E., Pavón-Domínguez, P., 2010b. Exploring the relationship between nitrogen dioxide and ground-level ozone by applying the joint multifractal analysis. Environmental Monitoring and Assessment 167, 675–684. https://doi.org/10.1007/s10661-009-1083-6

Kouvarakis, G., Tsigaridis, K., Kanakidou, M., Mihalopoulos, N., 2000. Temporal variations of surface regional background ozone over Crete Island in the southeast Mediterranean. Journal of Geophysical Research: Atmospheres 105, 4399–4407. https://doi.org/10.1029/1999JD900984

Lacasa, L., Luque, B., Ballesteros, F., Luque, J., Nuño, J.C., 2008. From time series to complex networks: The visibility graph. Proceedings of the National Academy of Sciences 105, 4972–4975.

https://doi.org/10.1073/pnas.0709247105

Lacasa, L., Luque, B., Luque, J., Nuño, J.C., 2009. The visibility graph: A new method for estimating the Hurst exponent of fractional Brownian motion. EPL (Europhysics Letters) 86, 30001.
https://doi.org/10.1209/0295-5075/86/30001

Lacasa, L., Toral, R., 2010. Description of stochastic and chaotic series using visibility graphs. Physical Review E 82.
https://doi.org/10.1103/PhysRevE.82.036120

Mali, P., Manna, S.K., Mukhopadhyay, A., Haldar, P.K., Singh, G., 2018. Multifractal analysis of multiparticle emission data in the framework of visibility graph and sandbox algorithm. Physica A: Statistical Mechanics and its Applications 493, 253–266.
https://doi.org/10.1016/j.physa.2017.10.015

Mandelbrot, B.B., 1982. The fractal geometry of nature. W.H. Freeman, San Francisco.

Pavon-Dominguez, P., Jiménez-Hornero, F.J., Gutierrez de Rave, E., 2013. Multifractal analysis of ground–level ozone concentrations at urban, suburban and rural background monitoring sites in Southwestern Iberian Peninsula. Atmospheric Pollution Research 4, 229–237.
https://doi.org/10.5094/APR.2013.024

Ribas, À., Peñuelas, J., 2004. Temporal patterns of surface ozone levels in different habitats of the North Western Mediterranean basin. Atmospheric Environment 38, 985–992. https://doi.org/10.1016/j.atmosenv.2003.10.045

Sillman, S., 1999. The relation between ozone, NOx and hydrocarbons in urban and polluted rural environments. Atmospheric Environment 33, 1821–1845. https://doi.org/10.1016/S1352-2310(98)00345-8

Tél, T., Fülöp, Á., Vicsek, T., 1989. Determination of fractal dimensions for geometrical multifractals. Physica A: Statistical Mechanics and its Applications 159, 155–166. https://doi.org/10.1016/0378-4371(89)90563-3

Telesca, L., Colangelo, G., Lapenna, V., Macchiato, M., 2004. Fluctuation dynamics in geoelectrical data: an investigation by using multifractal detrended fluctuation analysis. Physics Letters A 332, 398–404. https://doi.org/10.1016/j.physleta.2004.10.011

Trainer, M., Parrish, D.D., Goldan, P.D., Roberts, J., Fehsenfeld, F.C., 2000. Review of observation-based analysis of the regional factors influencing ozone concentrations. Atmospheric Environment 17.

Veneziano, D., Moglen, G.E., Bras, R.L., 1995. Multifractal analysis: Pitfalls of standard

procedures and alternatives. Phys. Rev. E 52, 1387–1398. https://doi.org/10.1103/PhysRevE.52.1387

Vicsek, T., Family, F., Meakin, P., 1990. Multifractal Geometry of Diffusion-Limited Aggregates. Europhysics Letters (EPL) 12, 217–222. https://doi.org/10.1209/0295-5075/12/3/005

Zeleke, T.B., Si, B.C., 2006. Characterizing scale-dependent spatial relationships between soil properties using multifractal techniques. Geoderma 134, 440–452. https://doi.org/10.1016/j.geoderma.2006.03.013

## 7. <u>FIGURES</u>

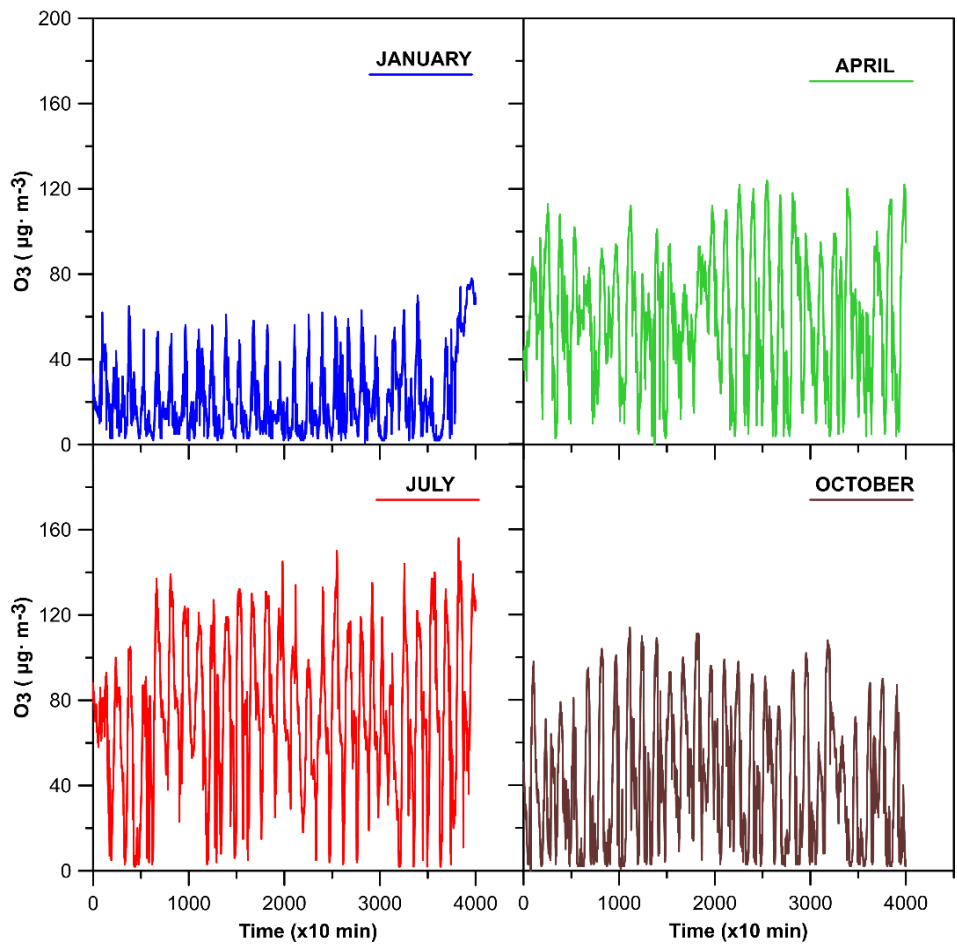

*Figure 1: Ozone time series for the different analyzed months.*

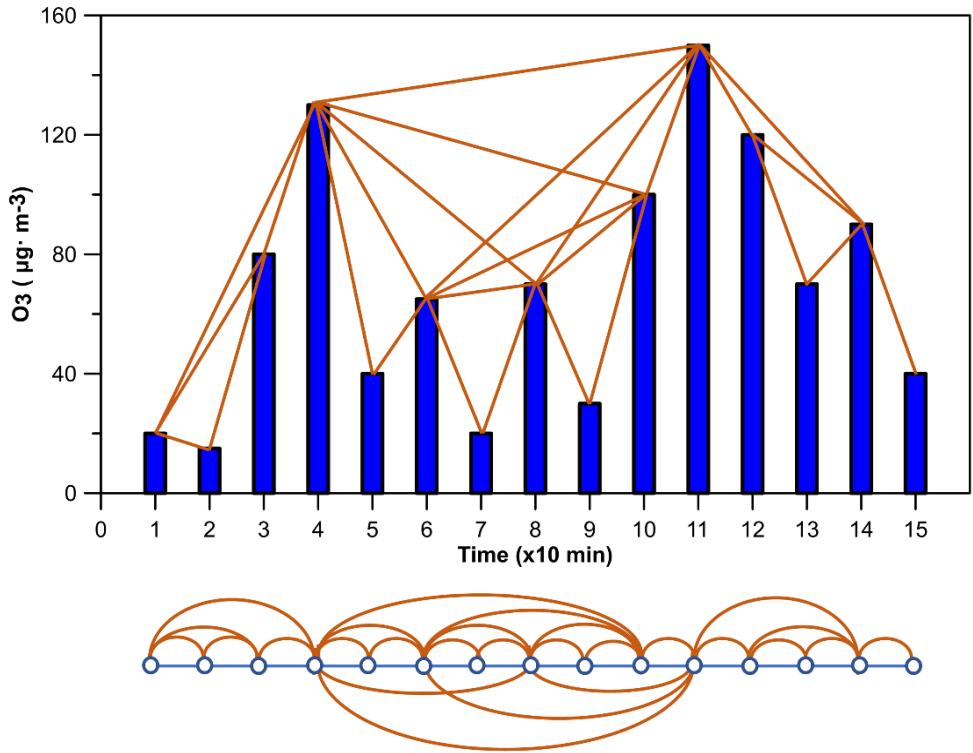

*Figure 2: Sample time series transformed into a complex network through the visibility algorithm. Below, all the connections are shown in a more visual way.*

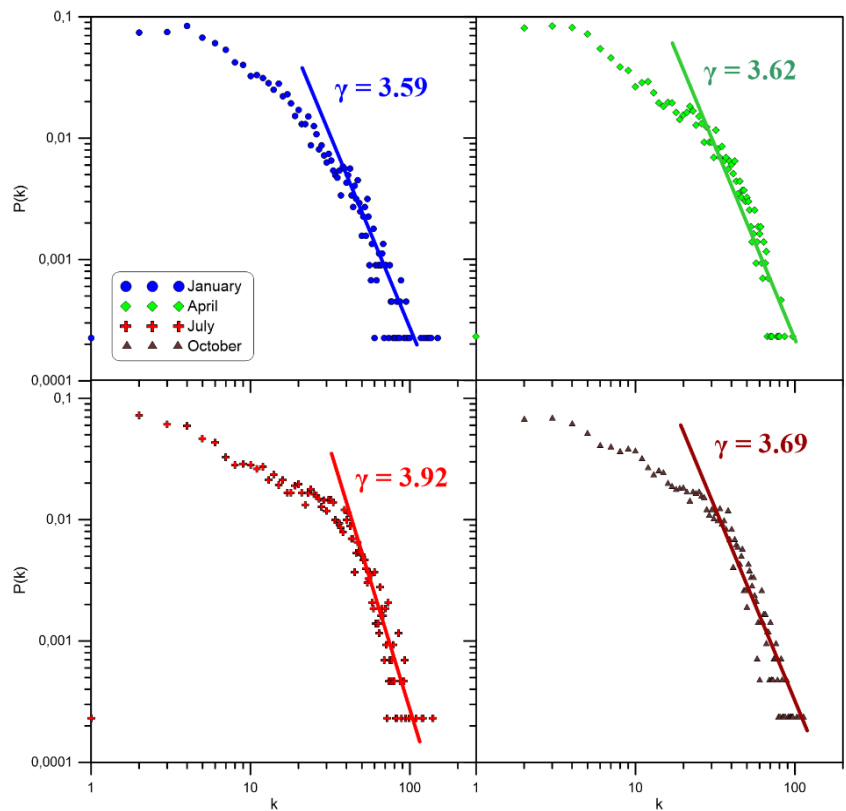

*Figure 3: Degree distribution of the visibility graph from each month in logarithmic scale. As it can be observed, the tail of the distribution shows a scale-free behavior because it can be fitted by a power law.*

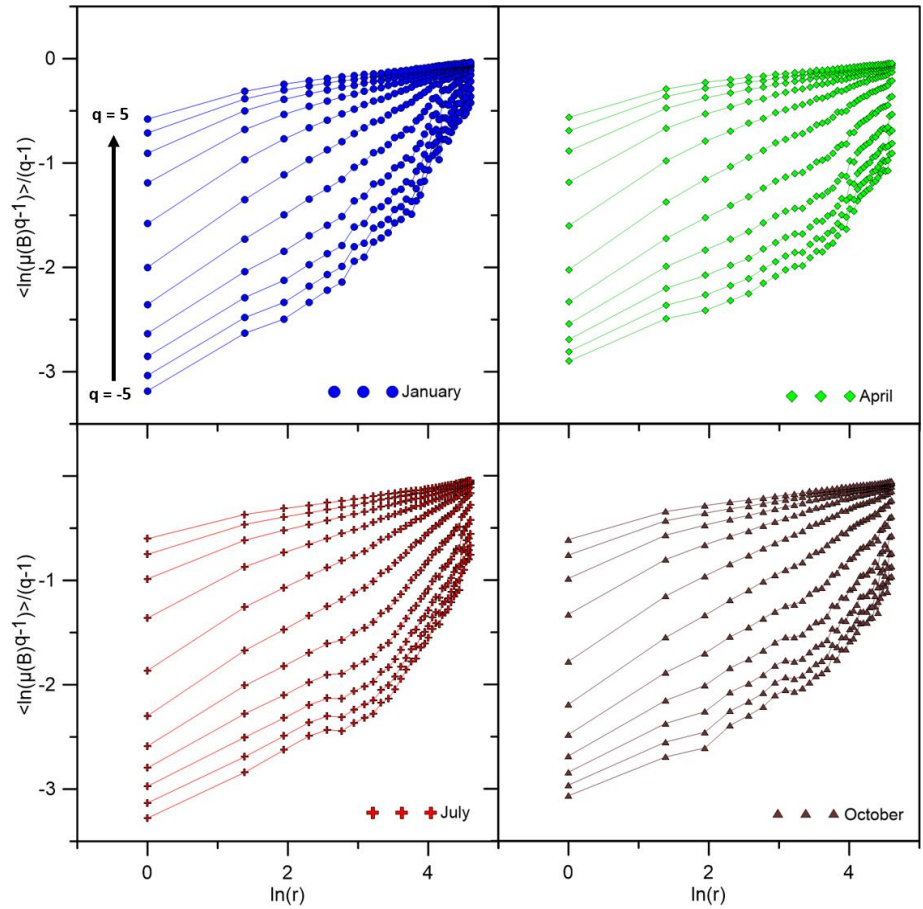

*Figure 4: Curves of the partition function against ln(r) obtained after applying the sandbox algorithm.*

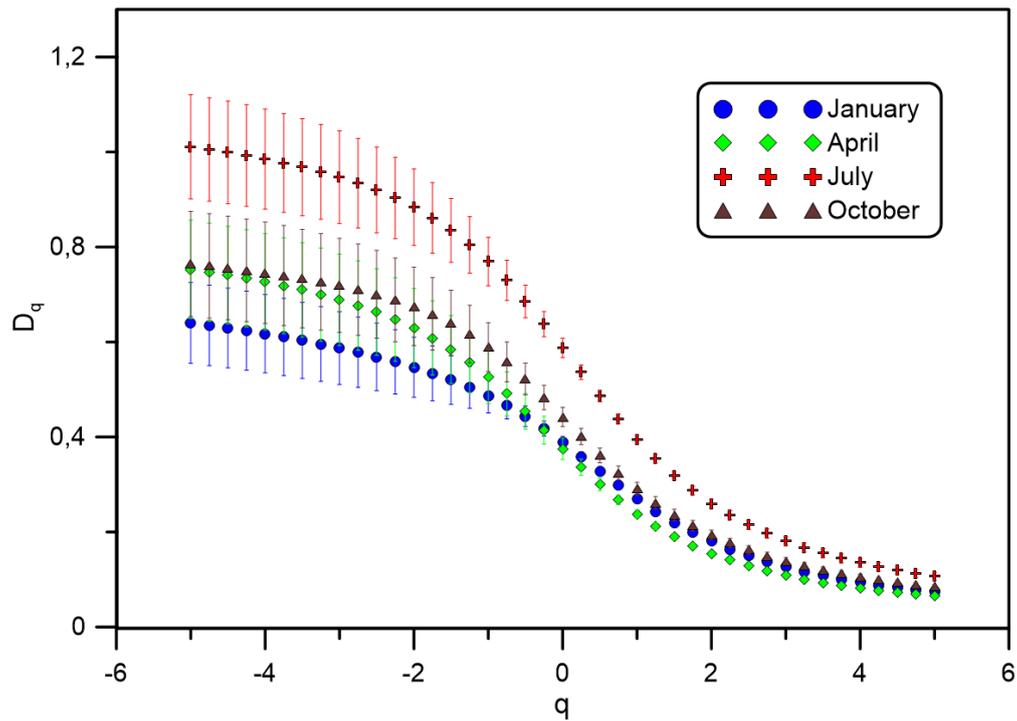

*Figure 5: Rényi dimensions for each month.*

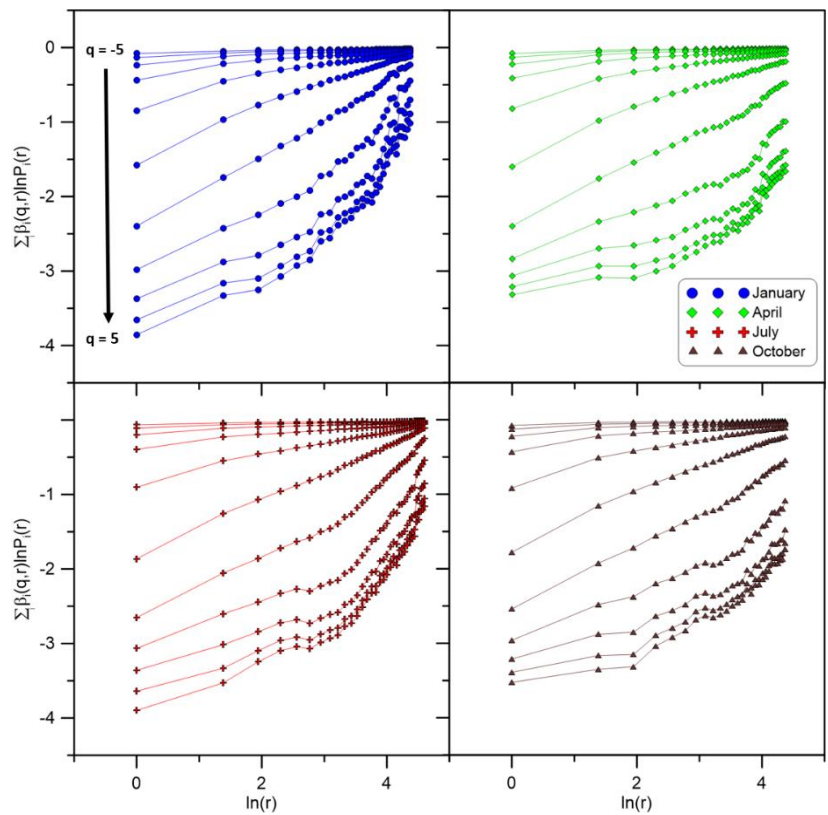

*Figure 6: Curves from where α is computed by regression.*

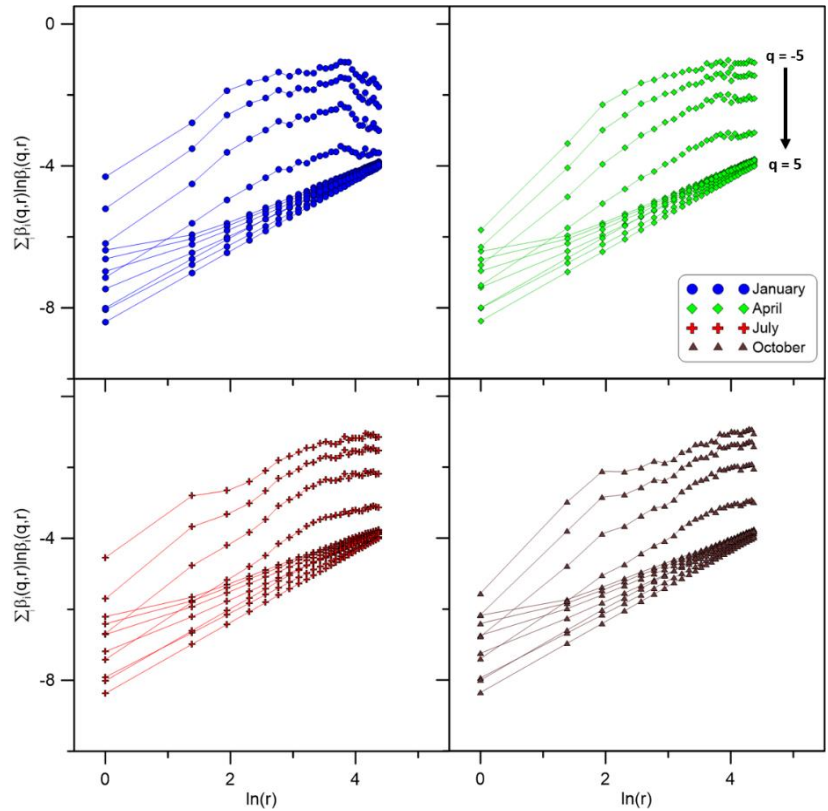

*Figure 7: Curves from where f(α) is computed using linear regressions.*

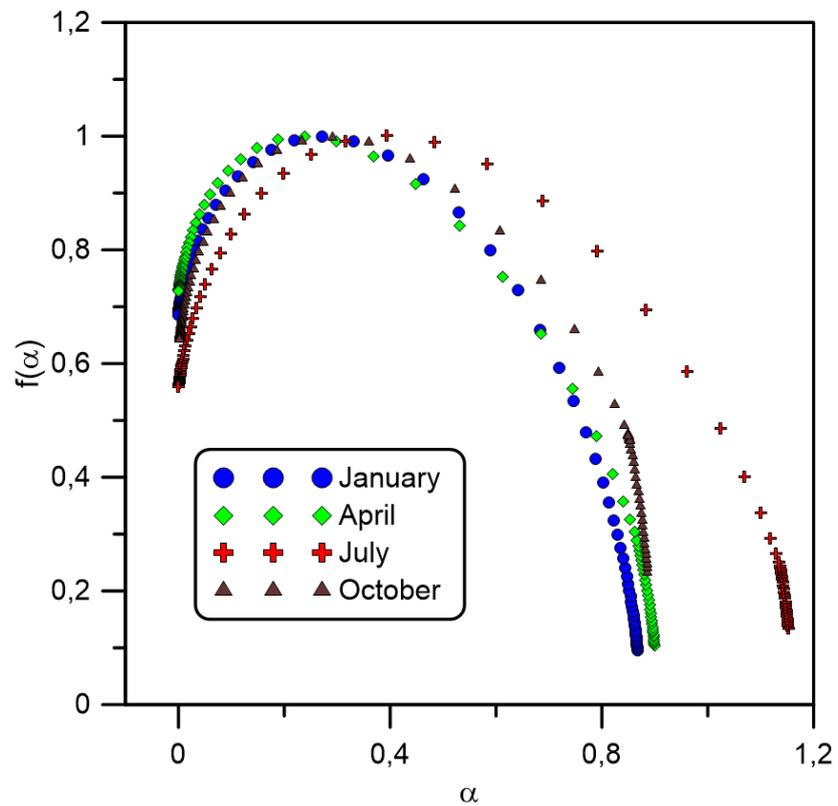

*Figure 8: α-spectra of all the months.*

8. <u>TABLES</u>

*Table 1: Relevant values for each month.*

| MONTH | Average Direct Radiation (MJ/m$^2$) | $\gamma$ | $\bar{k}$ | $\sigma_k$ | $D_0$ | $D_0 - D_1$ | $D_0 - D_2$ | $\Delta D_q$ | $\alpha_0$ | $W$ |
|---|---|---|---|---|---|---|---|---|---|---|
| January | 310.2 | 3.59 | 14.78 | 15.00 | 0.389±0.012 | 0.12±0.03 | 0.21±0.02 | 0.56±0.10 | 0.291 | 0.779 |
| April | 577.8 | 3.62 | 15.12 | 13.98 | 0.37±0.03 | 0.13±0.04 | 0.22±0.04 | 0.69±0.12 | 0.246 | 0.934 |
| July | 991.4 | 3.92 | 19.94 | 17.69 | 0.59±0.02 | 0.20±0.04 | 0.33±0.03 | 0.90±0.12 | 0.443 | 1.152 |
| October | 419.3 | 3.69 | 17.90 | 15.91 | 0.44±0.02 | 0.15±0.04 | 0.25±0.04 | 0.68±0.12 | 0.333 | 0.841 |

**HIGHLIGHTS**

- Ground-level ozone concentration time series have a multifractal behavior.

- Visibility graphs can be used to analyze multifractality of ozone time series.

- Many aspects of ozone dynamics can be observed through the degree distribution.

- This technique gives supplementary information within the study of complex signals.

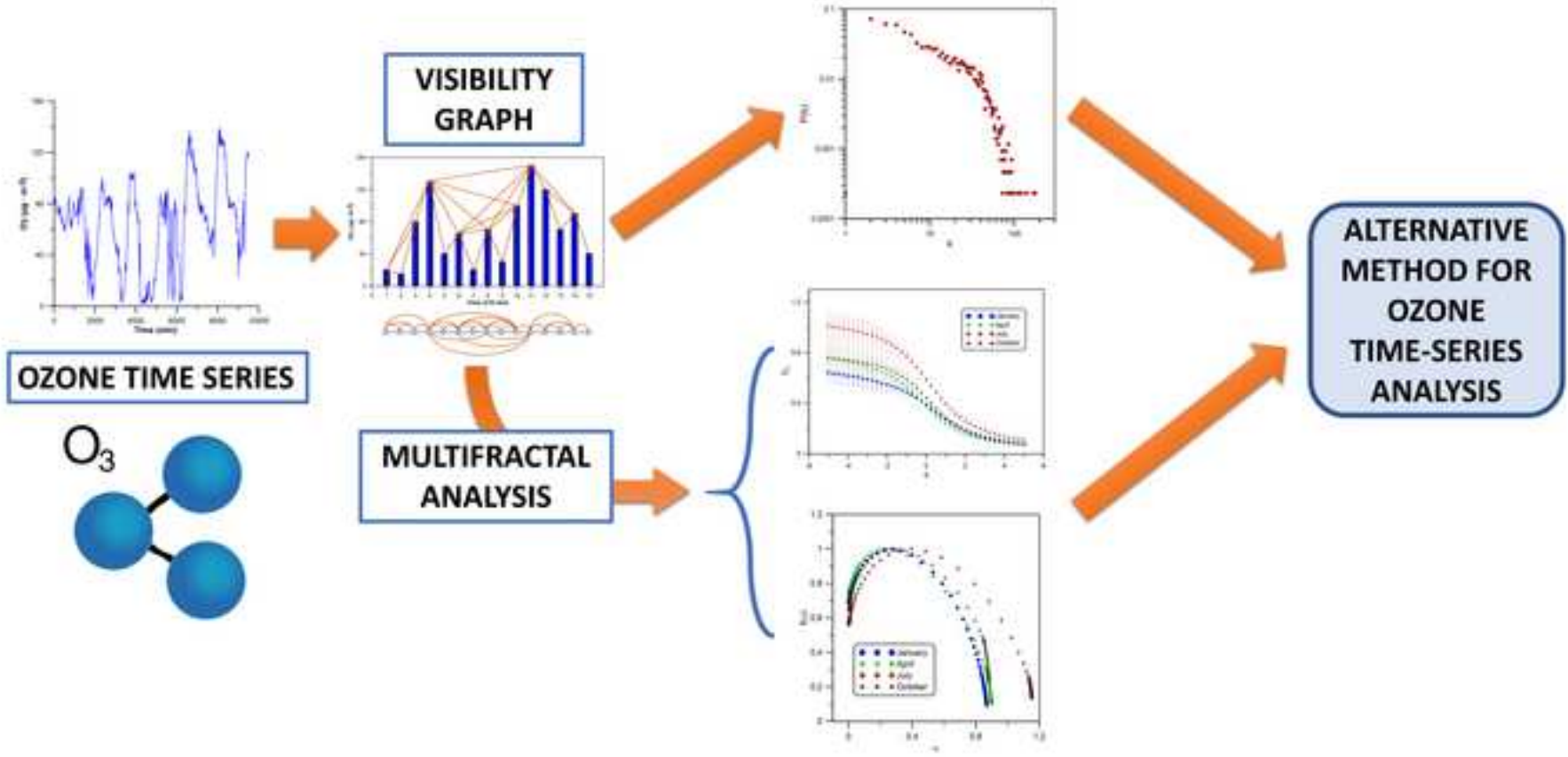